\documentclass[%
reprint,
amsmath,
amssymb,
aip,
jcp,            
nofootinbib
]{revtex4-2}

\usepackage{graphicx}
\usepackage{bm}
\usepackage{physics}
\usepackage{hyperref}

\hypersetup{
    colorlinks=true,
    linkcolor=blue,
    citecolor=blue,
    urlcolor=blue
}

\usepackage{lineno}

\usepackage{xcolor}
\usepackage[normalem]{ulem}
\usepackage[most]{tcolorbox}

\newif\iferbedit
\erbeditfalse   

\iferbedit
  \newcommand{\erbedit}[1]{\textcolor{blue}{#1}}
  \newcommand{\erbdel}[1]{\textcolor{red}{\sout{#1}}}
  
\else
  \newcommand{\erbedit}[1]{#1}
  \newcommand{\erbdel}[1]{}
  
\fi

\iferbedit
  \newenvironment{erbblock}
    {\color{blue}}
    {}
\else
  \newenvironment{erbblock}
    {}
    {}
\fi

\begin{document}


\title{Geometric Thermodynamics in Open Quantum Systems: Coherence, Curvature, and Work}
\author{Eric R. Bittner}
\affiliation{Department of Physics, University of Houston, Houston, Texas 77204, USA}

\date{\today}

\begin{abstract}
We formulate a geometric framework for quasistatic thermodynamics in open quantum systems by parameterizing the dynamics on a control manifold. In the quasistatic limit, the system follows a manifold of stationary states, and the work performed over a cycle is given by the flux of a curvature two-form, $W \sim \int \Omega$, defined by the parametric response of the stationary state, establishing an open-system analog of classical thermodynamic area laws. \erbedit{For thermal stationary states at fixed temperature, the curvature vanishes, reflecting the integrability of the work one-form.} Beyond this limit, nonequilibrium stationary states can retain coherence in the energy representation; using a fixed-basis Lindblad model, we show that this coherence reshapes the curvature, making it anisotropic and sign-changing, so that work depends sensitively on the placement and orientation of the cycle. Quantum coherence therefore partitions the control manifold into regions of opposite curvature, producing geometric cancellation of work and allowing the net work over a cycle to be reduced or reversed despite dissipative dynamics. Thermodynamic work thus emerges as a curvature flux whose structure is set by thermodynamic response in classical systems and by basis misalignment between the Hamiltonian eigenbasis and the environment-selected pointer basis in open quantum systems.
\end{abstract}
\maketitle

\section{Introduction}

Classical thermodynamics admits a geometric formulation in which equilibrium
states lie on Legendre submanifolds of a contact space, and reversible
processes correspond to trajectories constrained to these manifolds.
Thermodynamic work and heat are then naturally interpreted as geometric
quantities associated with closed cycles, such as areas enclosed in
conjugate planes. This structure, rooted in the invariance of the contact form under
Legendre transformations, provides a unified and coordinate-independent
description of equilibrium thermodynamics.

However, this geometric picture relies fundamentally on reversibility
and equilibrium.
In quantum and driven-dissipative systems, dynamics are governed not by
Hamiltonian flows on a fixed energy surface but by Liouvillian evolution
toward nonequilibrium steady states. In this setting, the notion of a
thermodynamic trajectory parametrized by time becomes ambiguous, and the
connection between cyclic processes and geometric quantities is no longer
evident.

In this work, we reformulate thermodynamic cycles as closed paths in a space
of control parameters $\lambda$, where the system is described at each point
by a Liouvillian $\mathcal{L}(\lambda)$ and relaxes to a corresponding
stationary state. This construction replaces time-parametrized dynamics with
a quasistatic progression along a manifold of steady states, restoring a
geometric interpretation of work and dissipation in a form that extends
naturally to open quantum systems.  
This construction restores a geometric notion of cyclic work while
allowing for dissipation and nonequilibrium steady-state structure.

Geometric approaches to thermodynamics have developed along several
complementary directions. In equilibrium, contact geometry provides a
natural framework in which thermodynamic variables define a contact
manifold and equilibrium states appear as Legendre
submanifolds~\cite{Hermann1973,Mrugala1991,Bravetti2017,Bravetti2015}.
Metric-based approaches due to Weinhold and Ruppeiner relate geometric
structure to equilibrium fluctuations and phase
behavior~\cite{Weinhold1975,Ruppeiner1995}. Extensions to nonequilibrium
systems include stochastic thermodynamics and fluctuation relations,
where work appears as a trajectory-dependent functional and satisfies
exact relations such as the Jarzynski equality and Crooks
theorem~\cite{Jarzynski1997,Crooks1999,Seifert2012}.

In the quantum regime, thermodynamic structure has been developed for open
systems governed by quantum dynamical semigroups and Lindblad
generators~\cite{Spohn1978,BreuerPetruccione,Alicki1979}, and more broadly
within the framework of quantum thermodynamics, including dynamical and
information-theoretic perspectives~\cite{Kosloff2013,Vinjanampathy2016,Kosloff2019,Esposito2009}. 
Considerable attention has also been given to driven quantum systems and 
quantum heat engines, where coherence and dissipation play competing 
roles~\cite{Scully2003,DeffnerLutz2011,Rahav2012,Dorfman2018,Camati2019,Klatzow2019}. 
More recently, geometric aspects of nonequilibrium quantum thermodynamics 
have been explored in the context of thermodynamic length, geometric 
optimization, and phase-like effects in driven open systems, including 
Berry-phase-induced transport and geometric response of thermal 
machines~\cite{Ren2010,Scandi2019,Abiuso2020,Brandner2020,TerrenAlonso2022,Bhandari2020}. 
These developments establish the foundations of quantum thermodynamics, but 
a direct geometric formulation of thermodynamic cycles—analogous to the 
classical area laws—has not been systematically developed.
A central goal of this work is to establish such a formulation directly
for open quantum systems, where the relevant states are nonequilibrium
stationary states of a Liouvillian generator rather than equilibrium
points on an energy surface.


For fixed control parameters, dissipative quantum dynamics admits
stationary states that play the role of nonequilibrium steady
states~\cite{BreuerPetruccione,Spohn1978}. By parameterizing the control
variables along a closed path and allowing the system to relax at each
step, one obtains a quasistatic cycle that is the direct analog of a
reversible thermodynamic process. This construction defines a manifold of
stationary states parameterized by the controls.


\erbedit{
Geometric aspects of driven open quantum systems have also been explored in the context of adiabatic pumping and full counting statistics, where cyclic variation of control parameters generates transport through Berry-phase mechanisms. In particular, Refs.~\citenum{Yadalam2016,Goswami2016} develop a geometric formulation in which pumped charge and higher cumulants are governed by a Berry connection defined from the eigenvectors of a (possibly counting-field–deformed) Liouvillian. These approaches establish that cyclic driving of nonequilibrium steady states can produce geometric contributions to transport observables.
}

\erbedit{
The geometric structure developed here is complementary in both construction and physical interpretation. Rather than arising from a gauge-dependent Berry connection associated with Liouvillian eigenmodes, the present formulation is defined directly in terms of steady-state thermodynamic response through the work one-form $A_W = \mathrm{Tr}(\rho_\star dH)$. The resulting curvature $\Omega_W = dA_W$ governs quasistatic work in control-parameter space and is expressed entirely in terms of observable expectation values, without requiring an eigenvector decomposition or counting-field formalism. This establishes a thermodynamic geometry associated with steady-state response, distinct from Berry-phase descriptions of transport, and provides a direct extension of classical cycle geometry to open quantum systems.}

Within this setting, we identify a quantum work one--form on the control
manifold and show that the work performed over a cycle is given by the
flux of an associated two--form. This provides a direct analog of the
classical area laws and establishes a geometric formulation of
thermodynamic cycles in open quantum systems.

A central feature of the quantum setting is the distinction between the
Hamiltonian eigenbasis and the pointer basis selected by the
environment~\cite{Zurek2003}. The resulting geometric structure is
controlled by the alignment between these bases. When they coincide, the
stationary state is diagonal in the energy representation and the
geometry reduces to a population-driven form. When they differ,
coherence persists in the stationary state and reshapes the geometric
response.

As a minimal example, we analyze a driven qubit with two control
parameters and obtain explicit expressions for the work one--form and
its curvature, providing a concrete setting in which coherence and
dissipation modify the underlying geometric structure.

Beyond formal considerations, there is a growing class of
driven-dissipative quantum systems in which such geometric structures may
be realized. Examples include exciton--polariton condensates, cavity QED
arrays, and driven solid-state systems, where steady states arise from a
balance of coherent driving and dissipation. In these systems, control
parameters such as pump intensity, detuning, or reservoir temperature can
be varied along closed paths while the system relaxes to a stationary
state at each point, realizing a quasistatic cycle on a manifold of
nonequilibrium steady states.

In particular, polariton condensates provide a natural platform in which
nonequilibrium steady states, coherence, and dissipation coexist. The
interplay between coherent light--matter coupling and reservoir-induced
relaxation leads to stationary states that are not thermal but nonetheless
well-defined. Cyclic modulation of system parameters—for example, through
pump shaping or cavity tuning—provides a route to defining and
experimentally accessing geometric work and curvature in direct analogy
with classical thermodynamic cycles.

From a theoretical perspective, such systems lie within the broader
framework of mesoscopic nonequilibrium thermodynamics, where dynamics can
be viewed as entropy-producing flows on a contact manifold. In this sense,
the geometric formulation developed here provides a bridge between
abstract thermodynamic structure and experimentally accessible
driven-dissipative quantum systems.
\section{Quasistatic Thermodynamics on the Liouvillian Control Manifold}

We consider a family of open quantum systems described by a
parameter-dependent Liouvillian
\begin{equation}
\dot{\rho} = \mathcal{L}(\lambda)[\rho],
\label{eq:L_general}
\end{equation}
where $\lambda = (\lambda^1, \lambda^2, \ldots)$ denotes externally
controlled parameters. These include both system parameters entering the
Hamiltonian $H(\lambda)$ and environmental variables such as temperature,
chemical potential, or system--bath coupling strengths. The control space
thus defines an extended manifold incorporating both system and bath
degrees of freedom and serves as the space of thermodynamic control
parameters.

\subsection{Stationary-state manifold and geometric cycles}

For each $\lambda$, dissipative dynamics drives the system to a stationary
state $\rho^\star(\lambda)$ satisfying
\begin{equation}
\mathcal{L}(\lambda)[\rho^\star(\lambda)] = 0.
\label{eq:ss_condition}
\end{equation}
We assume this state is unique and that the Liouvillian has a finite
spectral gap separating the zero eigenvalue from the rest of the spectrum,
ensuring relaxation on a well-defined timescale
\cite{Spohn1978,BreuerPetruccione}. Under these conditions, the stationary
state depends smoothly on the control parameters, as in adiabatic
response theory for open quantum systems~\cite{Avron2000,Sarandy2005}

The map
\[
\lambda \mapsto \rho^\star(\lambda)
\]
therefore defines a smooth manifold of stationary states embedded in the
space of density operators. This manifold provides the nonequilibrium
analog of the equilibrium thermodynamic manifold: it collects the states
selected by dissipative dynamics at fixed control parameters and defines
the space on which quasistatic processes are constructed.

A thermodynamic process is specified by a path $\lambda(s)$ in control
space, with a cycle corresponding to a closed loop $C$. In the quasistatic
limit, where the control parameters vary slowly compared to the
relaxation timescale, the system remains close to the stationary
manifold. In this regime, the dynamics reduces to motion along the
manifold itself, and the cycle becomes a geometric object determined
solely by the path in control space, in direct analogy with reversible
cycles in classical thermodynamics.

\paragraph{Liouvillian structure and quasistatic limit.}
We decompose the Liouvillian as
\begin{equation}
\mathcal{L}(\lambda) =
-\frac{i}{\hbar}[H(\lambda),\,\cdot\,] + \sum_k \mathcal{D}_k(\lambda),
\label{eq:L_decomp}
\end{equation}
where $\mathcal{D}_k$ are completely positive, trace-preserving
dissipators.

The quasistatic regime is defined by
\begin{equation}
\left| \frac{d\lambda}{dt} \right| \ll \Delta(\lambda),
\label{eq:qs_condition}
\end{equation}
where $\Delta(\lambda)$ is the Liouvillian spectral gap. In this limit,
the state admits an adiabatic expansion
\begin{equation}
\rho = \rho^\star + \delta\rho,
\end{equation}
with $\delta\rho$ small.
This expansion is the open-system analog of the adiabatic theorem: in the
quasistatic limit the system follows the instantaneous stationary state,
with corrections controlled by the inverse Liouvillian spectral gap
\cite{Sarandy2005,Avron2000}. This adiabatic response framework ensures
smooth parameter dependence and provides a controlled expansion governed
by the gap.

To leading order,
\erbedit{
\begin{equation}
\mathcal{L}(\lambda)[\delta\rho]
=
-\dot{\lambda}^i \partial_i \rho_\star(\lambda).
\label{eq:lin_eq}
\end{equation}
}
Defining the projector $\mathcal{P}[\cdot] = \rho^\star \mathrm{Tr}[\cdot]$
and $\mathcal{Q}=1-\mathcal{P}$, the inverse Liouvillian is well-defined on
the $\mathcal{Q}$ subspace, yielding
\begin{equation}
\delta\rho = -\mathcal{L}_\perp^{-1}
\left( \frac{d\rho^\star}{d\lambda} \right).
\label{eq:delta_rho_final}
\end{equation}
\erbedit{Here $\mathcal{L}_\perp^{-1}$ denotes the pseudoinverse of the Liouvillian restricted to the subspace orthogonal to the stationary state, defined by the projector $Q = 1 - P$, where $P[\rho] = \rho_\star \mathrm{Tr}(\rho)$.}

\paragraph{First law and geometric decomposition.}
The internal energy is
\begin{equation}
U = \mathrm{Tr}[\rho H],
\end{equation}
with differential
\begin{equation}
dU = \mathrm{Tr}[\rho\, dH] + \mathrm{Tr}[H\, d\rho].
\end{equation}
We identify
\begin{equation}
\delta W = \mathrm{Tr}[\rho\, dH], \qquad
\delta Q = \mathrm{Tr}[H\, d\rho],
\end{equation}
so that $dU = \delta W + \delta Q$
\footnote{We adopt the convention that $W>0$ corresponds to work performed on the
system. With this sign convention, the first law reads
$dU = \delta W + \delta Q$.}.

In the quasistatic limit,
\begin{equation}
\delta W =
\mathrm{Tr}[\rho^\star dH]
+
\mathrm{Tr}[\delta\rho\, dH].
\label{eq:work_split_final}
\end{equation}
The first term defines a one-form on control space,
\begin{equation}
\mathcal{A}_W = \mathrm{Tr}[\rho^\star dH],
\label{eq:AW_final}
\end{equation}
while the second encodes dissipative corrections.

\paragraph{Geometric work and curvature.}
The one-form $\mathcal{A}_W$ assigns to each displacement in control
space an infinitesimal work. For a closed cycle,
\begin{equation}
W_{\mathrm{rev}} = \oint_C \mathcal{A}_W,
\end{equation}
and by Stokes' theorem,
\begin{equation}
W_{\mathrm{rev}} = \iint_\Sigma \Omega_W,
\end{equation}
where
\begin{equation}
\Omega_W = d\mathcal{A}_W
= \mathrm{Tr}[d\rho^\star \wedge dH].
\label{eq:Omega_geom}
\end{equation}

Thus, work arises from the curvature associated with the parametric
response of the stationary state. Equivalently, nonzero cycle work
reflects the failure of the work one-form to be exact, or the
non-integrability of the map $\lambda \mapsto \rho^\star(\lambda)$, in
direct analogy with classical thermodynamic cycles.

Expanding
\begin{equation}
H = \sum_i X_i G_i,
\end{equation}
one obtains
\begin{equation}
\mathcal{A}_W = \sum_i \langle G_i \rangle_\star\, dX_i,
\end{equation}
so that expectation values act as generalized forces conjugate to the
control parameters. The curvature is
\begin{equation}
\Omega_W =
\sum_{i,j}
\left(
\partial_i \langle G_j \rangle_\star
-
\partial_j \langle G_i \rangle_\star
\right)
dX_i \wedge dX_j.
\end{equation}

This reproduces the classical structure of conjugate variables, but
arises here from the response of nonequilibrium stationary states rather
than from an underlying equilibrium equation of state.

\paragraph{Dissipation and thermodynamic metric.}
The dissipative contribution is
\begin{equation}
\delta W_{\mathrm{diss}} =
- \mathrm{Tr}\!\left[
\left(\mathcal{L}_\perp^{-1} \frac{d\rho^\star}{d\lambda}\right)
dH
\right].
\end{equation}
Writing $d\rho^\star = (\partial_i \rho^\star)d\lambda^i$, the total
dissipated work takes the quadratic form
\begin{equation}
W_{\mathrm{diss}} = \int dt\, g_{ij} \dot{\lambda}^i \dot{\lambda}^j,
\end{equation}
with
\begin{equation}
g_{ij} =
\mathrm{Tr}\!\left[
(\partial_i \rho^\star)\,
\mathcal{L}_\perp^{-1}
(\partial_j \rho^\star)
\right].
\end{equation}
This defines a positive semidefinite metric on control space, in direct
analogy with thermodynamic length and finite-time dissipation 
in classical nonequilibrium thermodynamics~\cite{Scandi2019,Abiuso2020,Brandner2020}.

The resulting geometry admits a natural decomposition into an
antisymmetric curvature $\Omega_W$, governing reversible work, and a
symmetric metric $g_{ij}$, governing dissipation.

\medskip
\paragraph{Thermal stationary states and contact structure.}
If the stationary state is thermal,
\begin{equation}
\rho_\beta =
\frac{e^{-\beta H}}{Z},
\end{equation}
then the von Neumann entropy satisfies
\begin{equation}
\delta Q = T\, dS_{\mathrm{vN}},
\end{equation}
and the First Law becomes
\begin{equation}
dU - T dS_{\mathrm{vN}} - \delta W = 0.
\end{equation}
This defines a one-form that vanishes on the stationary manifold, in
direct analogy with the contact condition of equilibrium thermodynamics.

In the quasistatic limit, the manifold of stationary states therefore
inherits a contact-geometric structure, and thermodynamic cycles appear
as closed curves on this manifold.
\section{Example: Qubit on an Extended Control Manifold}

We illustrate the geometric structure using a driven two-level system with Hamiltonian
\begin{equation}
H(\omega, g) = \frac{1}{2}(\omega \sigma_z + g \sigma_x),
\end{equation}
where $(\omega,g)$ serve as external control parameters. We couple the system to a thermal reservoir at temperature $T$, which we treat as an additional control variable. The resulting control manifold is therefore three-dimensional,
\begin{equation}
\lambda = (\omega, g, T).
\end{equation}

\subsection{Thermal stationary state: flat geometry}

We first consider the case in which the stationary state is thermal,
\begin{equation}
\rho_\star(\lambda) = \frac{e^{-\beta H(\omega,g)}}{Z(\lambda)}, \qquad \beta = \frac{1}{k_B T},
\end{equation}
as expected for dynamics that satisfy detailed balance.

The Hamiltonian eigenvalues are
\begin{equation}
E_\pm = \pm \frac{\epsilon}{2}, \qquad \epsilon = \sqrt{\omega^2 + g^2}.
\end{equation}
From these, we obtain the steady-state expectation values
\begin{equation}
\langle \sigma_z \rangle_\star =
- \frac{\omega}{\epsilon}
\tanh\!\left(\frac{\beta \epsilon}{2}\right), \qquad
\langle \sigma_x \rangle_\star =
- \frac{g}{\epsilon}
\tanh\!\left(\frac{\beta \epsilon}{2}\right).
\end{equation}

The work one-form follows directly,
\begin{equation}
\mathcal{A}_W =
\frac{1}{2}
\left(
\langle \sigma_z \rangle_\star \, d\omega
+
\langle \sigma_x \rangle_\star \, dg
\right).
\end{equation}
Substituting the expectation values gives
\begin{equation}
\mathcal{A}_W =
-\frac{1}{2}
\frac{\tanh(\beta \epsilon/2)}{\epsilon}
(\omega\, d\omega + g\, dg).
\end{equation}

At this point, the structure simplifies. Using
\begin{equation}
\omega\, d\omega + g\, dg = \epsilon\, d\epsilon,
\end{equation}
we obtain
\begin{equation}
\mathcal{A}_W =
-\frac{1}{2}
\tanh\!\left(\frac{\beta \epsilon}{2}\right)
\, d\epsilon.
\end{equation}

This form makes the geometry transparent. The work one-form depends only on the single scalar variable $\epsilon$, and is therefore exact:
\begin{equation}
\mathcal{A}_W = d\Phi(\epsilon), \qquad
\Phi(\epsilon) =
-\frac{1}{\beta}
\ln\!\left[2\cosh\!\left(\frac{\beta \epsilon}{2}\right)\right].
\end{equation}
Taking the exterior derivative yields
\begin{equation}
\Omega_W^{(\omega,g)} = d\mathcal{A}_W = 0.
\end{equation}

The structure of the one-form is now explicit. The dependence on the control parameters $(\omega,g)$ enters only through the single energy scale $\epsilon = \sqrt{\omega^2 + g^2}$. In effect, the mapping $(\omega,g) \mapsto \rho_\star$ collapses onto a one-dimensional manifold parameterized by $\epsilon$. As a consequence, the work one-form is integrable, and quasistatic work is path-independent in the $(\omega,g)$ plane at fixed temperature.
\erbedit{The associated curvature therefore vanishes identically, reflecting the integrability of the work one-form.}

\subsection{Curvature on the extended $(\omega,g,T)$ manifold}

Although the curvature vanishes in the $(\omega,g)$ subspace, the work one-form retains an explicit dependence on $\beta = 1/k_{B}T$. When we include temperature as a control parameter, the geometry becomes nontrivial.

Taking the exterior derivative on the full manifold yields
\begin{equation}
\Omega_W =
-\frac{\epsilon}{4}
\operatorname{sech}^2\!\left(\frac{\beta \epsilon}{2}\right)
\, d\beta \wedge d\epsilon.
\end{equation}
Expressed in terms of temperature,
\begin{equation}
\Omega_W =
\frac{\epsilon}{4 T^2}
\operatorname{sech}^2\!\left(\frac{\beta \epsilon}{2}\right)
\, dT \wedge d\epsilon.
\end{equation}

This curvature reflects the fact that temperature modifies the stationary state rather than the Hamiltonian itself. The geometric response, therefore, emerges only when the control protocol explores both energy and bath directions.

\subsection{Temperature-modulated cycles}

We now consider cyclic protocols on the extended manifold,
\begin{equation}
\omega(\theta) = \omega_0 + a \cos\theta, \qquad
g(\theta) = g_0 + b \sin\theta,
\end{equation}
with a temperature modulation
\begin{equation}
T(\theta) = T_0 + \Delta T \cos(\theta + \phi).
\end{equation}

The work performed over one cycle is
\begin{equation}
W_{\mathrm{cyc}} =
\frac{1}{2} \int_0^{2\pi}
\left(
\langle \sigma_z \rangle_\star \frac{d\omega}{d\theta}
+
\langle \sigma_x \rangle_\star \frac{dg}{d\theta}
\right)
d\theta,
\end{equation}
where the expectation values depend parametrically on $T(\theta)$.

Expanding about $T_0$ gives
\begin{equation}
W_{\mathrm{cyc}}(\phi)
\approx
W_0 + A \cos(\phi + \delta),
\end{equation}
showing that the net work depends on the relative phase between the geometric loop and the temperature modulation.

\medskip
\noindent
This example makes the central point explicit. Environmental parameters do not contribute directly to the work one-form. Instead, they reshape the stationary state, thereby modifying the geometric structure. Nontrivial curvature—and hence geometric work—emerges only when the control protocol probes multiple independent directions in parameter space.

\begin{erbblock}
\subsection{Generalization}
We now show that a flat 
geometry is a general feature of thermal mixed states. Let
\begin{equation}
\rho_\beta(\lambda)=\frac{e^{-\beta H(\lambda)}}{Z(\lambda)},
\qquad
Z(\lambda)=\Tr e^{-\beta H(\lambda)},
\end{equation}
be a Gibbs state at fixed inverse temperature $\beta$. Then the work one-form
\begin{equation}
A_i=\Tr\!\left(\rho_\beta\,\partial_i H\right)
\end{equation}
is exact:
\begin{equation}
A_i=\partial_i F_{\rm th}(\lambda),
\qquad
F_{\rm th}(\lambda)=-\frac{1}{\beta}\ln Z(\lambda).
\end{equation}
Hence the associated curvature vanishes identically,
\begin{equation}
\Omega_W=dA=0,
\qquad
F_{ij}=\partial_i A_j-\partial_j A_i=0.
\end{equation}

\noindent
The result follows directly from the definition of $A_i$. Differentiating the partition function gives
\begin{equation}
\partial_i \ln Z
=
-\beta\,\Tr\!\left(\rho_\beta\,\partial_i H\right),
\end{equation}
so that
\begin{equation}
A_i=-\frac{1}{\beta}\partial_i \ln Z=\partial_i F_{\rm th}(\lambda).
\end{equation}
Thus the work one-form is exact, $A=dF_{\rm th}$, and the curvature vanishes identically, $\Omega_W=dA=0$.

\medskip
\noindent
This result holds for any Gibbs family at fixed temperature and is not restricted to the two-level example. Thermal stationary states therefore define a geometrically flat reference manifold: the steady-state response is generated by a scalar potential, namely the free energy.

Nontrivial curvature requires either an extended control space that includes bath parameters such as $T$, or a departure from equilibrium steady-state structure. In particular, if $\beta$ is treated as a control parameter, the extended manifold $(\lambda,\beta)$ can support nonzero mixed components such as $\Omega_{\beta i}\neq 0$.

This flatness admits a simple geometric interpretation in the Bloch representation. For a Gibbs state, the steady-state Bloch vector is aligned with the Hamiltonian, $\mathbf{r} \propto \mathbf{h}$, so that their parametric variations remain locally parallel, $\partial_i \mathbf{r} \parallel \partial_i \mathbf{h}$. The antisymmetric combination in Eq.~(\ref{eq:bloch_curvature}) therefore vanishes identically, $F_{ij}=0$.

Geometric curvature thus requires a breakdown of this alignment. When the steady state is no longer determined solely by the instantaneous Hamiltonian, the response of the state and the deformation of the Hamiltonian become misaligned, generating a nonzero curvature and geometric work.

Although the curvature becomes nonzero upon extending the control manifold to include temperature, its structure remains highly constrained. For a Gibbs state, the steady-state density matrix depends only on the ratio $\beta \epsilon$, where $\epsilon=\sqrt{\omega^2+g^2}$ is the instantaneous energy scale. As a result, the curvature on the extended $(\epsilon,T)$ manifold collapses to a universal scaling form
\[
\Omega_W^{\rm th}(\epsilon,T)
=
\frac{\epsilon}{4T^2}\sech^2\!\left(\frac{\epsilon}{2T}\right),
\]
or equivalently $T\Omega_W^{\rm th} = \tfrac{x}{2}\sech^2 x$ with $x=\epsilon/(2T)$. This demonstrates that the thermal curvature is effectively one-dimensional, governed entirely by the ratio $\epsilon/T$, and reflects a population susceptibility rather than a genuinely multidimensional geometric structure. The curvature is positive definite, peaked near $\epsilon\sim T$, and vanishes both in the small-gap and large-gap limits as shown in 
Fig.~\ref{fig:coherent_vs_thermal_curvature}.

This effective dimensional reduction is a direct consequence of the fact that the Gibbs state depends only on a single energy scale. In more complex interacting systems, where multiple energy scales and collective fluctuations enter, this constraint is lifted and the curvature can acquire a genuinely multidimensional structure. This behavior is explored in a companion study of interacting lattice models, where the geometric curvature develops extended regions of enhanced response associated with critical fluctuations\cite{bittner2026thermodynamiccurvaturewidomridge}.

\end{erbblock}

\section{Quantum Geometry from Basis Misalignment}
\erbedit{In the preceding example the stationary state is diagonal in the
instantaneous energy eigenbasis of the Hamiltonian. The geometric
response is therefore entirely population-driven and depends only on the
energy scale $\epsilon$, revealing an effectively one-dimensional
structure.}

To move beyond this limit, we consider stationary density matrices
$\rho^\star$ satisfying $\mathcal{L}[\rho^\star]=0$. The Hamiltonian
defines an energy eigenbasis, while the environment selects a pointer
basis through the system--bath coupling~\cite{Zurek2003}. The stationary
condition fixes $\rho^\star$ in this pointer basis, where it is diagonal.
If the pointer basis differs from the eigenbasis of $H$, then
$[\rho^\star,H]\neq 0$. In the energy representation, this misalignment
is coherence.

Geometry therefore reflects basis misalignment. When the two bases
coincide, the stationary state is diagonal in the generator of motion
and the response is purely population-driven. When they differ,
coherence enters the stationary state and reshapes the geometric
response.

We retain the qubit Hamiltonian
\begin{equation}
H(\omega,g)=\frac{1}{2}(\omega\sigma_z+g\sigma_x),
\end{equation}
but define dissipation in a fixed laboratory basis,
\begin{equation}
\dot{\rho}
=
-\frac{i}{\hbar}[H,\rho]
+\gamma_\downarrow \mathcal{D}_{\sigma_-}[\rho]
+\gamma_\uparrow \mathcal{D}_{\sigma_+}[\rho],
\label{eq:coherent_lindblad}
\end{equation}
with
\begin{equation}
\mathcal{D}_{L}[\rho]
=
L\rho L^\dagger
-\frac12\{L^\dagger L,\rho\}.
\end{equation}
The dissipators are diagonal in the $\sigma_z$ basis, which therefore
acts as a pointer basis. Because $H$ contains both $\sigma_z$ and
$\sigma_x$, its eigenbasis is rotated relative to this pointer basis
whenever $g\neq 0$. The stationary state is set by the competition
between coherent precession and dissipative relaxation, and in general
does not commute with $H$.

Writing
\begin{equation}
\rho=\frac12\left(I+x\sigma_x+y\sigma_y+z\sigma_z\right),
\end{equation}
the steady-state solution is
\begin{align}
x^\star &= \frac{2s\,\omega g}{\gamma D}, \\
y^\star &= -\frac{s\,g}{D}, \\
z^\star &= \frac{s\left(4\omega^2+\gamma^2\right)}{2\gamma D},
\end{align}
with
\begin{equation}
D = 2\omega^2 + g^2 + \frac{\gamma^2}{2},\qquad
\gamma=\gamma_\downarrow+\gamma_\uparrow,\quad
s=\gamma_\downarrow-\gamma_\uparrow.
\end{equation}

The steady-state Bloch vector is not aligned with the effective
Hamiltonian field $(g,0,\omega)$. This misalignment reflects the
competition between coherent evolution and dissipative selection of the
pointer basis and appears as coherence in the energy representation.

The generalized forces therefore measure coherence, not just population.
This coherence enters directly in the work one-form,
\begin{equation}
\mathcal{A}_W
=
\frac12\left(z^\star\,d\omega + x^\star\,dg\right),
\end{equation}
and generates curvature
\begin{equation}
\Omega_W^{\rm coh}
=p
\frac{\,g\left(g^2+\gamma^2\right)}
{\left(2\omega^2+g^2+\gamma^2/2\right)^2}
\, d\omega\wedge dg,
\end{equation}
with anisotropy and sign.
The dimensionless parameter $p$ measures the thermal bias of the bath,
interpolating between $p=1$ (zero temperature) and $p=0$ (infinite
temperature).

\begin{figure*}[t]
\centering
\includegraphics[width=0.95\linewidth]{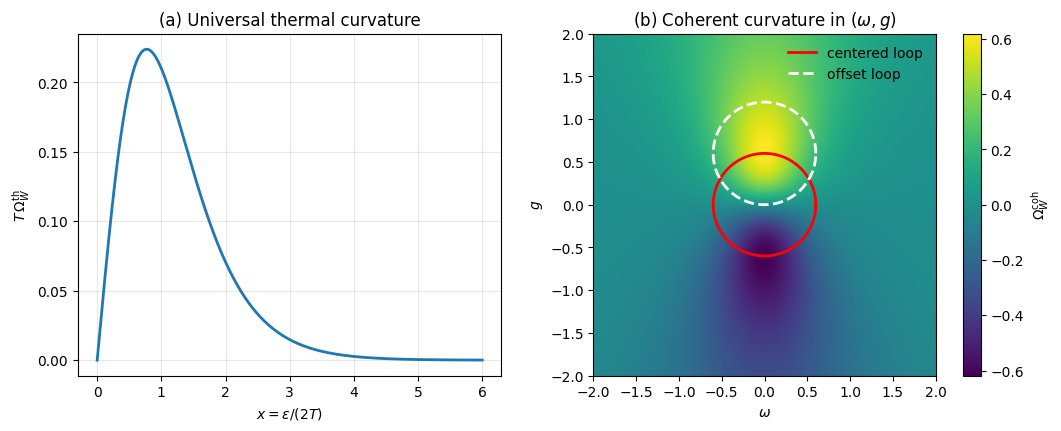}
\caption{
\erbedit{
Comparison of geometric curvature in thermal and coherent regimes. 
(a) Thermal curvature on the extended Gibbs-state control manifold, shown as the universal scaling function $T\Omega_W^{\rm th}$ versus $x=\epsilon/(2T)$. At fixed temperature the curvature vanishes identically in the $(\omega,g)$ plane; nonzero curvature arises only when the protocol explores both energy and bath directions. The curvature is positive definite and peaked near $\epsilon\sim T$, reflecting the regime of maximal sensitivity of the Gibbs populations to temperature. 
(b) Coherent curvature $\Omega_W^{\rm coh}$ for the fixed-basis Lindblad model, evaluated in the $(\omega,g)$ plane at fixed temperature. The curvature is anisotropic and changes sign due to the mismatch between the Hamiltonian eigenbasis and the environment-selected pointer basis. The dashed white loop is symmetric about $g=0$ and encloses regions of opposite curvature, leading to cancellation of the net work, while the solid red loop is displaced and samples predominantly one sign, producing finite work.}
}
\label{fig:coherent_vs_thermal_curvature}
\end{figure*}

\erbedit{
Figure~\ref{fig:coherent_vs_thermal_curvature} shows the resulting
geometry. In the thermal case, the curvature vanishes identically in
the $(\omega,g)$ plane at fixed temperature and, when extended to include
temperature, collapses to a positive, universal function of $\epsilon/T$,
reflecting an effectively one-dimensional response. In contrast, the
coherent case exhibits a curvature that changes sign across the control
manifold. Work becomes a balance between positive and negative
contributions.
}

Two loops illustrate the effect. A loop centered at $g=0$ samples
positive and negative curvature nearly symmetrically, and the net work
cancels. A loop displaced into the $g>0$ sector samples predominantly one
sign and produces finite work. The response is therefore controlled not
only by area, but by where the cycle sits in the curvature landscape.
\erbedit{When the stationary state is diagonal in the Hamiltonian basis, the curvature on the extended control manifold is positive definite and entirely population-driven.}

Thermodynamic curvature therefore emerges from basis misalignment between
the system and its environment. When the Hamiltonian eigenbasis and the
pointer basis coincide, the geometry is effectively classical. When they
differ, coherence reshapes the curvature and renders the work sensitive
to loop placement and orientation.

More general dissipative models interpolate between these limits by
rotating the effective pointer basis toward the Hamiltonian eigenbasis.
The fixed-basis model isolates the mechanism in its simplest form.

\subsection{Geometric response in the Bloch representation}

The preceding analysis identifies quasistatic work as the flux of a curvature two-form over the control manifold. This curvature reflects the non-integrability of the work one-form and signals that the steady-state response cannot be derived from a scalar thermodynamic potential, in contrast to equilibrium systems.

Geometric approaches to nonequilibrium thermodynamics also introduce symmetric, metric-like structures that characterize dissipation. In slow transformations between steady states, entropy production is governed by quadratic forms in the control velocities, defining a Riemannian metric that quantifies the cost of driving. Thermodynamic response therefore separates into two complementary geometric sectors: a symmetric sector associated with dissipation and an antisymmetric sector associated with geometric work.
Similar decompositions have been identified in recent work on geometric response theory, where symmetric and antisymmetric contributions arise from the underlying structure of the steady-state manifold and its parametric dependence~\cite{Bettmann:2025aa,Lacerda:2025aa}.

The work one-form takes the form
\begin{equation}
A_i
=
\mathrm{Tr}\!\left(\rho_{\mathrm{ss}} \, \partial_{\lambda_i} H\right)
=
\frac{1}{2}\,\mathbf{r} \cdot \partial_{\lambda_i} \mathbf{h},
\end{equation}
where $\rho_{\mathrm{ss}} = (I + \mathbf{r}\cdot\boldsymbol{\sigma})/2$ is the steady-state density matrix defined in terms of the Bloch vector $\mathbf{r}$, and
\begin{equation}
H = \frac{1}{2}\,\mathbf{h}(\lambda)\cdot\boldsymbol{\sigma}.
\end{equation}

To place state and control variations on equal footing, we introduce the dimensionless control vector
\begin{equation}
\tilde{\mathbf{h}} = \beta \mathbf{h},
\end{equation}
which naturally appears in the Gibbs state and defines the thermodynamically relevant control geometry.

The associated curvature is
\begin{equation}
F_{ij}
=
\partial_i A_j - \partial_j A_i
=
\frac{1}{2}
\left[
\partial_i \mathbf{r} \cdot \partial_j \mathbf{h}
-
\partial_j \mathbf{r} \cdot \partial_i \mathbf{h}
\right],
\label{eq:bloch_curvature}
\end{equation}
which depends only on the physical Hamiltonian vector $\mathbf{h}$ and sets the scale of geometric work.

Equation~(\ref{eq:bloch_curvature}) makes the mechanism explicit. Geometric work arises from an antisymmetric coupling between the parametric response of the steady-state Bloch vector and the deformation of the Hamiltonian. Curvature vanishes whenever $\partial_i \mathbf{r}$ and $\partial_i \mathbf{h}$ remain locally aligned, as occurs for thermal states or in the strong dephasing limit. Nonzero curvature, therefore, requires a genuine misalignment between state adaptation and Hamiltonian deformation, induced by coherence or competing dissipative structures.

This motivates a unified geometric description in which variations of the state and Hamiltonian enter on equal footing. Define the Hermitian tensor
\begin{equation}
\mathcal{T}_{ij}
=
\partial_i \mathbf{r} \cdot \partial_j \mathbf{r}
+
\partial_i \tilde{\mathbf{h}} \cdot \partial_j \tilde{\mathbf{h}}
+
i\left(
\partial_i \mathbf{r} \cdot \partial_j \tilde{\mathbf{h}}
-
\partial_j \mathbf{r} \cdot \partial_i \tilde{\mathbf{h}}
\right),
\label{eq:response_tensor}
\end{equation}
which is dimensionless and defined entirely on the control manifold.  
The antisymmetric part of $\mathcal{T}_{ij}$ reproduces the curvature governing geometric work up to an overall thermodynamic scale,
\begin{equation}
F_{ij} = \frac{1}{2\beta}\,\mathrm{Im}\,\mathcal{T}_{ij},
\end{equation}
while the symmetric part defines a quadratic form,
\begin{equation}
g_{ij} \equiv \mathrm{Re}\,\mathcal{T}_{ij}
=
\partial_i \mathbf{r} \cdot \partial_j \mathbf{r}
+
\partial_i \tilde{\mathbf{h}} \cdot \partial_j \tilde{\mathbf{h}}.
\end{equation}

The first contribution,
\begin{equation}
g^{(r)}_{ij} = \partial_i \mathbf{r} \cdot \partial_j \mathbf{r},
\end{equation}
is closely related to the quantum Fisher information metric. For a qubit,
\begin{equation}
g^{\mathrm{QFI}}_{ij}
=
\partial_i \mathbf{r}\cdot \partial_j \mathbf{r}
+
\frac{(\mathbf{r}\cdot \partial_i \mathbf{r})(\mathbf{r}\cdot \partial_j \mathbf{r})}{1 - |\mathbf{r}|^2},
\end{equation}
so that $g^{(r)}_{ij}$ captures the leading, transverse contribution to the quantum Fisher information. The second term accounts for variations in the magnitude of the Bloch vector and becomes important for strongly polarized states.

Equivalently, the Bures metric, which provides the natural information-geometric structure on quantum state space, is given by
\begin{equation}
g^{\mathrm{Bures}}_{ij}
=
\frac{1}{4}
\left[
\partial_i \mathbf{r} \cdot \partial_j \mathbf{r}
+
\frac{(\mathbf{r} \cdot \partial_i \mathbf{r})(\mathbf{r} \cdot \partial_j \mathbf{r})}{1 - |\mathbf{r}|^2}
\right],
\end{equation}
and is proportional to the quantum Fisher information metric that governs the statistical distinguishability of nearby states \cite{BraunsteinCaves1994,Paris2009}. The associated distance is derived from quantum fidelity \cite{Bures1969,Uhlmann1976}.

In this form, $\partial_i \mathbf{r} \cdot \partial_j \mathbf{r}$ measures the transverse sensitivity of the steady state, while the full quantum Fisher information incorporates longitudinal variations associated with changes in state purity. In the weak-polarization limit, $g^{\mathrm{Bures}}_{ij} \approx g^{(r)}_{ij}/4$, and $g^{(r)}_{ij}$ provides a direct measure of the information-geometric response of the steady state.

The second contribution,
\begin{equation}
g^{(h)}_{ij} = \partial_i \tilde{\mathbf{h}} \cdot \partial_j \tilde{\mathbf{h}},
\end{equation}
describes the geometry of the control manifold itself, independent of the state. It measures how rapidly the Hamiltonian direction changes in parameter space and provides a purely kinematic contribution to the cost of driving.
Taken together, $\mathcal{T}_{ij}$ defines a response tensor that unifies dissipative and geometric aspects of thermodynamic response. Its symmetric part $g_{ij}$ sets the local cost of driving through information and control geometry, while its antisymmetric part generates geometric work, with the physical curvature obtained as $F_{ij} = \frac{1}{2\beta}\,\mathrm{Im}\,\mathcal{T}_{ij}$.

This decomposition parallels the quantum geometric tensor, which separates into a metric and a Berry curvature. Here, however, the structure is defined on a manifold of nonequilibrium steady states and explicitly couples the state's response to the deformation of the Hamiltonian. The tensor $\mathcal{T}_{ij}$ is dimensionless, while the curvature governing work inherits its physical scale through $\beta^{-1}$. 
Geometric work thus emerges as a thermodynamically scaled antisymmetric component of a response tensor, directly measuring the failure of state response and Hamiltonian deformation to commute under parameter variation.

\section{Coherence-Induced Cancellation of Geometric Work}
As shown in Fig.~\ref{fig:coherent_vs_thermal_curvature},
quantum coherence in the quasi-stationary state partitions the control manifold into regions of positive and negative curvature.
The local
curvature is diagnostic; the physically relevant quantity is the work
accumulated over a closed path. This motivates a global measure that
compares coherent and population-driven geometric response.

For a cycle $C$ enclosing a surface $\Sigma(C)$, we define the geometric
work-reduction factor
\begin{equation}
\eta_{\rm geom}[C]
=
\frac{W_{\rm coh}[C]}{W_{\rm pop}[C]}
=
\frac{\iint_{\Sigma(C)} \Omega_W^{\rm coh}}
     {\iint_{\Sigma(C)} \Omega_W^{\rm pop}}.
\label{eq:eta_geom_def}
\end{equation}
Here $W_{\rm pop}$ is evaluated from the positive-definite curvature on the extended control manifold (e.g., including temperature), which provides a natural population-driven reference,  while $W_{\rm coh}$ is computed from the
full coherent curvature, which can change sign across the control
manifold. The ratio $\eta_{\rm geom}$ isolates the effect of coherence
on geometric work: $\eta_{\rm geom}<1$ corresponds to a reduction of the
work input due to cancellation between positive and negative curvature
regions, $\eta_{\rm geom}=0$ indicates complete cancellation, and
$\eta_{\rm geom}<0$ corresponds to a reversal of the net geometric work
due to dominance of opposite-sign curvature.
Thus, $\eta_{\rm geom}$ quantifies how quantum coherence redistributes
geometric work by inducing cancellation of curvature flux across the
control manifold.

\paragraph{Symmetric cycles.}
For any cycle symmetric under $g\rightarrow -g$, such as the centered
loop in Fig.~\ref{fig:coherent_vs_thermal_curvature}, the coherent
curvature is antisymmetric,
\[
\Omega_W^{\rm coh}(\omega,g)=-\Omega_W^{\rm coh}(\omega,-g),
\]
and therefore integrates to zero,
\begin{equation}
W_{\rm coh}[C]=0,
\qquad
\eta_{\rm geom}[C]=0.
\end{equation}
Coherence  enforces exact cancellation between regions of opposite
curvature, suppressing the net work geometrically even in the presence
of dissipation.

\paragraph{Displaced cycles.}
For cycles displaced away from symmetry points, such as the offset loop
in Fig.~\ref{fig:coherent_vs_thermal_curvature}, the antisymmetry of
$\Omega_W^{\rm coh}$ no longer enforces cancellation. The net work is
set by how the cycle samples the curvature landscape. In this regime,
the geometric work-reduction factor depends not only on the size of the
cycle, but on its placement and orientation in control space.
Coherence now introduces a new degree of control: shifting the cycle across
the curvature landscape tunes the net work.

\paragraph{Relation to local response.}
For sufficiently small cycles centered at $(\omega_0,g_0)$, the work is
set by the local curvature, and Eq.~(\ref{eq:eta_geom_def}) reduces to
\begin{equation}
\eta_{\rm geom}[C]
\approx
\frac{\Omega_W^{\rm coh}(\omega_0,g_0)}
     {\Omega_W^{\rm pop}(\omega_0,g_0)}.
\end{equation}
This local approximation is useful, but the geometric effect is
intrinsically nonlocal, arising from the integrated curvature over the
cycle.

If the dissipative dynamics satisfies detailed balance, the population
asymmetry parameter becomes
\begin{equation}
p = \tanh\!\left(\frac{\beta\epsilon}{2}\right),
\qquad
\epsilon=\sqrt{\omega^2+g^2}.
\end{equation}
In this case, both the coherent and population-driven curvatures inherit
a common thermal dependence, while the variation of $\epsilon$ along the
cycle renders the weighting of curvature regions path-dependent even at
fixed temperature.

For small cycles centered at $(\omega_0,g_0)$, the geometric
work-reduction factor simplifies to
\begin{equation}
\eta_{\rm geom}(\omega_0,g_0)
\approx
\frac{
2\,g_0(g_0^2+\gamma^2)\,
\sinh(\beta\epsilon_0)
}{
\beta\left(2\omega_0^2+g_0^2+\gamma^2/2\right)^2
},
\end{equation}
with 
\begin{equation}
\epsilon_0=\sqrt{\omega_0^2+g_0^2}.
\end{equation}
This expression makes explicit that the coherent geometric response is set by a
competition between thermal suppression of the population-driven
curvature, encoded in $\mathrm{sech}^2(\beta\epsilon_0/2)$, and
coherence-induced amplification, encoded in $\sinh(\beta\epsilon_0)$.
As the population-driven baseline becomes increasingly localized at low
temperature, the relative importance of coherence correspondingly grows.

\paragraph{Geometric interpretation and implications.}
The quantity $\eta_{\rm geom}$ has a direct geometric meaning. In the
population-driven case, the curvature is positive and the work is set by
the enclosed area weighted by a positive density. In the coherent case,
the curvature defines a signed measure on control space, and the work is
given by the net signed flux through the cycle. Coherence therefore acts
as a geometric interference mechanism, producing cancellation between
different regions of control space. The thermodynamic response is no
longer determined solely by the size of the cycle, but by how it is
embedded within the curvature landscape.

We emphasize that $\eta_{\rm geom}$ is not a thermodynamic efficiency
\emph{per se}, but rather a geometric measure of the relative magnitude
of coherent and population-driven contributions to the work. This
identifies a new mechanism for controlling thermodynamic response in open
quantum systems: by engineering the alignment between the Hamiltonian
eigenbasis and the environment-selected pointer basis, one can shape the
curvature and thereby tune the geometric work. Cycles can be designed to
minimize work input through controlled cancellation, or enhance it by
selectively sampling regions of a given sign.
\section{Fluctuating Work and Geometric Structure}

The geometric formulation developed above describes quasistatic cycles
as deterministic trajectories on the control manifold. In realistic
settings, however, work is an integrated quantity defined along a
trajectory.
%
\erbedit{
Fluctuations in the work arise from stochasticity in the system trajectory induced by environmental interactions. In the geometric formulation, this can be represented equivalently as stochastic sampling of paths on the control manifold.
}
The work therefore becomes a random
functional of the path sampled on the control manifold.
For driven processes, the Jarzynski equality relates these fluctuations
to the free-energy difference,
\begin{align}
\left\langle e^{-\beta W} \right\rangle = e^{-\beta \Delta F},
\end{align}
where the average is taken over realizations of the protocol
\cite{Jarzynski1997,Crooks1999,Seifert2012}.

Within the present framework, the work performed along a control protocol
$\lambda(t)$ is a geometric functional defined by a line integral of a
connection one-form $\mathcal A_W$ over a path in control space,
\begin{align}
W[\lambda(t)] = \int_{\lambda(t)} \mathcal A_W .
\end{align}
For a fixed protocol $\lambda(t)$, this expression is entirely
deterministic. Fluctuations in the work arise from stochasticity in the
control trajectory itself, which we regard as a realization drawn from a
probability measure $\mathcal{P}[\lambda(t)]$ over paths in parameter
space. The work therefore becomes a random variable induced by the
distribution of trajectories.

\begin{widetext}
With this interpretation, the Jarzynski equality is naturally written as
an expectation value over control-space paths,
\begin{align}
\left\langle
\exp\!\left[-\beta \int_{\lambda(t)} \mathcal A_W \right]
\right\rangle
=
\int \mathcal{D}\lambda \, \mathcal{P}[\lambda(t)]
\exp\!\left[-\beta \int_{\lambda(t)} \mathcal A_W \right]
=
e^{-\beta \Delta F},
\label{eq:Jarzynski_connection}
\end{align}
where $\Delta F$ is the equilibrium free energy difference between the
endpoints of the protocol. 

For closed quasistatic cycles, or for open protocols completed by a
reference path on the stationary manifold, Stokes' theorem converts the
line integral into a surface integral of the curvature 2-form
$\Omega_W = d\mathcal A_W$,
\begin{align}
W[\lambda(t)] = \iint_{\Sigma(\lambda)} \Omega_W ,
\end{align}
where $\Sigma(\lambda)$ is any surface bounded by the trajectory
$\lambda(t)$ together with the chosen reference path. In this form, the
work depends only on the geometric flux of $\Omega_W$ through the surface.

The Jarzynski equality can then be recast as an average over such
surfaces,
\begin{align}
\left\langle
\exp\!\left[-\beta \iint_{\Sigma(\lambda)} \Omega_W \right]
\right\rangle
=
e^{-\beta \Delta F}.
\label{eq:Jarzynski_curvature}
\end{align}


Stochastic fluctuations can be incorporated at the level of the control
parameters by promoting $\lambda(t)$ to a diffusion process on the
control manifold,
\begin{align}
d\lambda^i = v^i(\lambda)\,dt + \sigma^{ij}(\lambda)\,dW_j,
\label{eq:control_sde}
\end{align}
where $v^i(\lambda)$ is the drift, $\sigma^{ij}(\lambda)$ the noise
matrix, and $dW_j$ are independent Wiener increments. In this setting,
the work becomes a stochastic geometric functional,
\begin{align}
W = \int \mathcal A_i(\lambda)\circ d\lambda^i,
\label{eq:stochastic_work}
\end{align}
defined in the Stratonovich sense to preserve covariance under changes
of control coordinates.

The resulting dynamics lifts the geometric work functional to a diffusion
process on the extended space $(\lambda,W)$. The corresponding
Fokker--Planck equation for the joint distribution $P(\lambda,W,t)$ is
derived in Appendix~\ref{app:stochastic_geometry}. In compact form, the
evolution is governed by
\begin{align}
\partial_t P
=
-\partial_i(v^i P)
-\partial_W\!\left(b_W P\right)
+\frac12 \partial_i\partial_j(D^{ij}P)
+\partial_i\partial_W\!\left(D^{ij}\mathcal A_j P\right)
+\frac12 \partial_W^2\!\left(\mathcal A_i D^{ij}\mathcal A_j P\right),
\label{eq:FP_main}
\end{align}
where $D^{ij}=\sigma^{ik}\sigma^{jk}$ and
\begin{align}
b_W = \mathcal A_i v^i + \frac12 D^{ij}\partial_i \mathcal A_j.
\end{align}
This equation provides a direct route to computing the full statistics of
geometric work from stochastic control dynamics. In particular, diffusion of the control parameters induces diffusion of
the accumulated work through the geometric coupling encoded in
$\mathcal A_i$, so that work fluctuations arise from stochastic sampling
of the connection on the control manifold.
\end{widetext}

This analysis makes it explicit that thermodynamic fluctuations are
encoded in a statistical ensemble of paths (or equivalently surfaces) in
control space, while the thermodynamic response is governed by the
underlying geometric structure. The work is determined by a gauge-like
connection $\mathcal A_W$, and its fluctuations arise from stochastic
sampling of trajectories, providing a direct link between
nonequilibrium thermodynamics and geometry on the control manifold.

Equation~(\ref{eq:Jarzynski_curvature}) gives a geometric interpretation
of fluctuation relations: the Jarzynski average becomes an ensemble
average over fluctuating curvature fluxes. In the thermal,
population-driven regime, where $\Omega_W$ is positive, the work
distribution is dominated by contributions of a fixed sign and the
deterministic area law emerges as the mean response. In the coherent
regime, by contrast, $\Omega_W$ changes sign. The work functional then
samples regions of opposite curvature, so that different realizations
partially cancel through the signed flux.
In this regime, fluctuations do not simply broaden the work
distribution; \textit{they probe a control manifold with intrinsically signed
curvature.} Coherence therefore modifies both the mean work and its
higher moments through a geometric mechanism. From this perspective, the
Jarzynski equality acquires a natural interpretation as an average over
fluctuating geometric actions.   In this sense, quantum coherence enters fluctuation thermodynamics as a source of
signed geometric action.

\section{Conclusions}

We have developed a geometric formulation of quasistatic thermodynamics
for open quantum systems based on the structure of the Liouvillian
control manifold. By parameterizing the dynamics in terms of externally
controlled variables and allowing the system to relax to its stationary
state at each point, we obtain a nonequilibrium analog of the equilibrium
thermodynamic manifold. On this manifold, reversible work defines a
one-form whose exterior derivative yields a curvature two-form, and the
work performed over a cycle is given by the flux of this curvature.

\erbedit{
For thermal stationary states, the geometry closely parallels the
classical case in that it remains integrable in structure. The curvature
vanishes identically in the $(\omega,g)$ plane at fixed temperature and,
when extended to include temperature, reduces to a positive, universal
function of the ratio $\epsilon/T$. As a result, work accumulation is
governed by a scalar response associated with population sensitivity,
rather than a genuinely multidimensional curvature field. Environmental
parameters such as temperature do not enter directly in the work one-form,
but reshape the curvature landscape, controlling how geometric work is
distributed across energy scales and sampled along a cycle.
In this sense, the Gibbs state provides a minimal integrable limit of thermodynamic geometry, in which curvature, when present, remains constrained to a single effective direction in control space.
}

\erbedit{Extending beyond thermal stationary states, we show that genuinely
quantum geometric effects arise when the stationary density matrix is not
diagonal in the energy eigenbasis. In this regime, the competition between
coherent dynamics and dissipative relaxation produces steady states that
retain coherence in the energy representation. The resulting curvature is
anisotropic and sign-changing, so that work depends not only on the size
of the cycle, but also on its placement and orientation in control space.}


\emph{We show that quantum coherence partitions the control manifold into
regions of opposite curvature, producing geometric cancellation of work
over cyclic protocols.}
As a result, coherence can reduce or even reverse the net work performed
over a cycle, despite fully dissipative dynamics. This identifies a
genuinely quantum mechanism—absent in classical thermodynamics—in which
interference in the stationary state reshapes the geometric response.

This leads to a simple geometric principle: \emph{thermodynamic curvature is
set by the relation between the Hamiltonian eigenbasis and the pointer
basis selected by the environment.} When these bases coincide, the
geometry is classical and population-driven. When they differ, coherence
reshapes the curvature and introduces directional and sign-sensitive
structure. Thermodynamic curvature emerges as a direct manifestation of
basis misalignment.

The framework admits direct realization in molecular and condensed-phase
systems in which effective Hamiltonian parameters are controlled by
external fields, nuclear configurations, or structured electromagnetic
environments. In particular, molecular polariton systems provide a
natural platform, where a reduced two-state subspace is formed by a
molecular excitation and a cavity photon. Cyclic modulation of detuning,
light--matter coupling, or reservoir conditions defines trajectories in
control space along which the system relaxes to nonequilibrium steady
states, enabling direct probing of geometric work and curvature.
More generally, the geometric response can be controlled by tuning both
the Hamiltonian and the system--bath coupling. By adjusting detuning,
hybridization, and environmental interactions, one can design cycles that
selectively sample regions of positive or negative curvature, thereby
controlling work input, extraction, and flow. 
\erbedit{This perspective elevates thermodynamic response from a passive property
to a tunable geometric feature of the manifold, linking microscopic
dynamics to macroscopic work and connecting geometric thermodynamics with
driven steady-state physics and quantum control.}
In this sense, the framework provides a route to
controlling thermodynamic response in open quantum systems and connects
geometric thermodynamics with driven steady-state physics and quantum
control.

\vspace{0.5cm}

\section*{Data Availability Statement}
All data generated or analyzed during this study are included in this manuscript.

\begin{acknowledgments}
This work at the University of Houston was supported by the National Science Foundation under CHE-2404788 and the Robert A. Welch Foundation (E-1337).
\end{acknowledgments}

\section*{Author Contributions}
The author developed the theoretical framework, performed all derivations,
and carried out the analysis presented in this work. Generative AI tools
were used during manuscript preparation to assist with drafting,
structural organization, and verification of intermediate algebraic
steps. All theoretical formulations, results, and physical
interpretations were independently developed and validated by the author.

\section*{Conflicts of Interest}
The author declares no competing financial or non-financial interests.

\appendix
\begin{widetext}

\section{Stochastic Control Dynamics and Geometric Work}
\label{app:stochastic_geometry}
In this appendix we formulate the stochastic dynamics of the control
parameters and provide the technical details required to derive the
Fokker--Planck equation for the geometric work functional.

\subsection{Stochastic dynamics on the control manifold}

We treat the control variables $\lambda^i$ as a diffusion process on the
control manifold,
\begin{equation}
d\lambda^i = v^i(\lambda)\,dt + \sigma^{ij}(\lambda)\,dW_j,
\end{equation}
where $v^i(\lambda)$ is the drift vector, $\sigma^{ij}(\lambda)$ is the
noise matrix, and $dW_j$ are independent Wiener increments satisfying
\begin{equation}
\langle dW_i\rangle = 0,
\qquad
\langle dW_i\,dW_j\rangle = \delta_{ij}\,dt.
\end{equation}
Here $\lambda^i$ denotes coordinates on the control manifold, and
$\mathcal A_i$ is the associated work one-form.
It is convenient to define the diffusion tensor
\begin{equation}
D^{ij}(\lambda) = \sigma^{ik}(\lambda)\sigma^{jk}(\lambda),
\end{equation}
where repeated indices are summed over.

\subsection{Geometric work as a stochastic functional}

The work is defined as a geometric line integral of the connection
one-form $\mathcal A_i(\lambda)$,
\begin{equation}
W_t = \int_0^t \mathcal A_i(\lambda(\tau)) \circ d\lambda^i(\tau),
\end{equation}
where $\circ$ denotes the Stratonovich integral. This choice is natural
geometrically, as it preserves the standard chain rule under coordinate
transformations on the control manifold.

To derive the associated stochastic dynamics, we convert to Ito form
using the standard Stratonovich-to-Ito relation,
\begin{equation}
\mathcal A_i(\lambda)\circ d\lambda^i
=
\mathcal A_i(\lambda)\,d\lambda^i
+\frac12 D^{ij}(\lambda)\,\partial_i \mathcal A_j(\lambda)\,dt.
\end{equation}
Substituting the stochastic differential equation for $d\lambda^i$
then yields
\begin{equation}
dW =
\left[
\mathcal A_i(\lambda) v^i(\lambda)
+\frac12 D^{ij}(\lambda)\partial_i \mathcal A_j(\lambda)
\right] dt
+
\mathcal A_i(\lambda)\, \sigma^{ij}(\lambda) dW_j.
\end{equation}

Thus the pair $(\lambda,W)$ evolves as a Markov diffusion process in the
extended space,
\begin{align}
d\lambda^i &= v^i(\lambda)\,dt + \sigma^{ij}(\lambda)\, dW_j, \\
dW &= b_W(\lambda)\,dt + c_j(\lambda)\,dW_j,
\end{align}
with coefficients
\begin{align}
b_W(\lambda) &= \mathcal A_i(\lambda) v^i(\lambda)
+\frac12 D^{ij}(\lambda)\partial_i \mathcal A_j(\lambda), \\
c_j(\lambda) &= \mathcal A_i(\lambda)\, \sigma^{ij}(\lambda).
\end{align}
This form makes explicit that stochasticity in the control variables
induces both drift and diffusion in the accumulated work through the
geometric coupling encoded in $\mathcal A_i$.

\subsection{Fokker--Planck equation in $(\lambda,W)$ space}

Let $P(\lambda,W,t)$ denote the joint probability density for the
control variables and accumulated work. The corresponding Fokker--Planck
equation is
\begin{align}
\partial_t P
&=
-\partial_i(v^i P)
-\partial_W(b_W P)
+\frac12 \partial_i \partial_j (D^{ij} P)
+\partial_i \partial_W (M^i P)
+\frac12 \partial_W^2 (N P),
\end{align}
where
\begin{align}
M^i(\lambda) &= D^{ij}(\lambda)\mathcal A_j(\lambda), \\
N(\lambda) &= \mathcal A_i(\lambda) D^{ij}(\lambda)\mathcal A_j(\lambda).
\end{align}

Substituting the explicit expressions yields
\begin{align}
\partial_t P
&=
-\partial_i(v^i P)
-\partial_W\!\left[
\left(\mathcal A_i v^i + \tfrac12 D^{ij}\partial_i \mathcal A_j\right) P
\right] \nonumber \\
&\quad
+\frac12 \partial_i \partial_j (D^{ij} P)
+\partial_i \partial_W \left(D^{ij} \mathcal A_j P\right)
+\frac12 \partial_W^2 \left(\mathcal A_i D^{ij} \mathcal A_j P\right).
\label{eq:FP_appendix}
\end{align}

The structure of this equation admits a clear geometric interpretation.
The first and third terms describe drift and diffusion of the control
variables on the manifold, while the second term governs the deterministic
drift of the accumulated work. The mixed derivative term,
\[
\partial_i \partial_W \left(D^{ij}\mathcal A_j P\right),
\]
encodes correlations between motion on the control manifold and work
fluctuations, and the final term,
\[
\frac12 \partial_W^2 \left(\mathcal A_i D^{ij} \mathcal A_j P\right),
\]
describes diffusion of work induced by stochastic wandering of the
control trajectory.

\subsection{Moment generating function and tilted generator}

It is convenient to introduce the moment generating function
\begin{equation}
Z(\lambda,\chi,t)
=
\int dW\, e^{-\chi W} P(\lambda,W,t).
\end{equation}
Multiplying the Fokker--Planck equation by $e^{-\chi W}$ and integrating
over $W$ yields a closed evolution equation,
\begin{equation}
\partial_t Z = \mathcal L_\chi Z,
\end{equation}
with tilted generator
\begin{align}
\mathcal L_\chi
&=
-\partial_i\!\left(v^i\,\cdot\right)
+\frac12 \partial_i \partial_j \!\left(D^{ij}\,\cdot\right)
+\chi \left(\mathcal A_i v^i + \tfrac12 D^{ij}\partial_i \mathcal A_j\right) \nonumber \\
&\quad
-\chi \,\partial_i \!\left(D^{ij}\mathcal A_j\,\cdot\right)
+\frac{\chi^2}{2} \mathcal A_i D^{ij} \mathcal A_j.
\end{align}
At $\chi=\beta$, this operator generates the Jarzynski average and
encodes the full work statistics within a tilted diffusion on the
control manifold.

\subsection{Isotropic diffusion}

For isotropic, constant noise with $D^{ij} = 2D\,\delta^{ij}$, the
coefficients simplify to
\begin{align}
b_W &= \mathcal A_i v^i + D\,\partial_i \mathcal A_i, \\
M^i &= 2D\,\mathcal A_i, \\
N &= 2D\,\mathcal A_i \mathcal A_i,
\end{align}
and the Fokker--Planck equation reduces to
\begin{align}
\partial_t P
&=
-\partial_i(v^i P)
-\partial_W\!\left[(\mathcal A_i v^i + D\,\partial_i \mathcal A_i) P\right]
+D\,\partial_i \partial_i P \nonumber \\
&\quad
+2D\,\partial_i \partial_W (\mathcal A_i P)
+D\,\partial_W^2 \left[(\mathcal A_i \mathcal A_i) P\right].
\end{align}

\subsection{Geometric interpretation}

The accumulation of work along a trajectory is governed by the connection
$\mathcal A_i$, while circulation around closed loops is determined by
the curvature
\begin{equation}
\Omega_{ij} = \partial_i \mathcal A_j - \partial_j \mathcal A_i,
\end{equation}
which is invariant under gauge transformations
$\mathcal A_i \rightarrow \mathcal A_i + \partial_i \phi$.
For open stochastic trajectories, work statistics depend on the choice of
connection and are therefore gauge-dependent, whereas for closed cycles
the gauge-invariant contribution is given by the curvature flux. In this
sense, stochastic work accumulation separates into connection-dependent
local contributions and curvature-controlled geometric contributions for
closed protocols. This provides the stochastic extension of the geometric formulation
developed in the main text.

\end{widetext}


\bibliographystyle{apsrev4-2}

\bibliography{geom_oqs_literature_v2}

@misc{bittner2026thermodynamiccurvaturewidomridge,
      title={Thermodynamic Curvature and the Widom Ridge in Interacting Spin Systems}, 
      author={Eric R. Bittner},
      year={2026},
      eprint={2604.16707},
      archivePrefix={arXiv},
      primaryClass={cond-mat.stat-mech},
      url={https://arxiv.org/abs/2604.16707}, 
}

@article{Yadalam2016,
  author = {Yadalam, H. K. and Harbola, U.},
  title = {Statistics of an adiabatic charge pump},
  journal = {Physical Review B},
  volume = {93},
  number = {3},
  pages = {035312},
  year = {2016},
  doi = {10.1103/PhysRevB.93.035312}
}

@article{Goswami2016,
  author = {Goswami, H. P. and Agarwalla, B. K. and Harbola, U.},
  title = {Geometric effects in nonequilibrium electron transfer statistics in adiabatically driven quantum junctions},
  journal = {Physical Review B},
  volume = {93},
  number = {19},
  pages = {195441},
  year = {2016},
  doi = {10.1103/PhysRevB.93.195441}
}

@misc{Bettmann:2025aa,
	archiveprefix = {arXiv},
	author = {Laetitia P. Bettmann and Artur M. Lacerda and Mark T. Mitchison and John Goold},
	eprint = {2512.05074},
	primaryclass = {quant-ph},
	title = {Thermodynamic universality across dissipative quantum phase transitions},
	url = {https://arxiv.org/abs/2512.05074},
	year = {2025}}

@article{Lacerda:2025aa,
	author = {Lacerda, Artur M. and Bettmann, Laetitia P. and Goold, John},
	doi = {10.1103/9f6l-d766},
	issue = {2},
	journal = {Phys. Rev. E},
	month = {Aug},
	numpages = {6},
	pages = {L022101},
	publisher = {American Physical Society},
	title = {Information geometry of transitions between quantum nonequilibrium steady states},
	url = {https://link.aps.org/doi/10.1103/9f6l-d766},
	volume = {112},
	year = {2025}}

@article{Paris2009,
	author = {Paris, Matteo G. A.},
	journal = {International Journal of Quantum Information},
	pages = {125--137},
	title = {Quantum estimation for quantum technology},
	volume = {7},
	year = {2009}}

@article{Bures1969,
	author = {Bures, Donald},
	journal = {Transactions of the American Mathematical Society},
	pages = {199--212},
	title = {An Extension of Kakutani's Theorem on Infinite Product Measures to the Tensor Product of Semifinite w*-Algebras},
	volume = {135},
	year = {1969}}

@article{Uhlmann1976,
	author = {Uhlmann, Armin},
	journal = {Reports on Mathematical Physics},
	pages = {273--279},
	title = {The "transition probability" in the state space of a *-algebra},
	volume = {9},
	year = {1976}}

@article{BraunsteinCaves1994,
	author = {Braunstein, Samuel L. and Caves, Carlton M.},
	journal = {Physical Review Letters},
	pages = {3439--3443},
	title = {Statistical distance and the geometry of quantum states},
	volume = {72},
	year = {1994}}

@article{Sarandy2005,
	author = {Sarandy, Marcelo S. and Lidar, Daniel A.},
	doi = {10.1103/PhysRevA.71.012331},
	journal = {Phys. Rev. A},
	pages = {012331},
	title = {Adiabatic approximation in open quantum systems},
	volume = {71},
	year = {2005}}

@article{Kosloff2013,
	author = {Kosloff, Ronnie},
	doi = {10.3390/e15062100},
	journal = {Entropy},
	number = {6},
	pages = {2100--2128},
	title = {Quantum Thermodynamics: A Dynamical Viewpoint},
	volume = {15},
	year = {2013}}

@article{Kosloff2019,
	author = {Kosloff, Ronnie},
	doi = {10.3390/e21050540},
	journal = {Entropy},
	number = {5},
	pages = {540},
	title = {Quantum Thermodynamics and Open-Systems Modeling},
	volume = {21},
	year = {2019}}

@article{Vinjanampathy2016,
	author = {Vinjanampathy, Sai and Anders, Janet},
	doi = {10.1080/00107514.2016.1201896},
	journal = {Contemp. Phys.},
	number = {4},
	pages = {545--579},
	title = {Quantum Thermodynamics},
	volume = {57},
	year = {2016}}

@article{Esposito2009,
	author = {Esposito, Massimiliano and Harbola, Upendra and Mukamel, Shaul},
	journal = {Rev. Mod. Phys.},
	pages = {1665--1702},
	title = {Nonequilibrium Fluctuations, Fluctuation Theorems, and Counting Statistics in Quantum Systems},
	volume = {81},
	year = {2009}}

@article{Spohn1978,
	author = {Spohn, Herbert},
	journal = {J. Math. Phys.},
	pages = {1227},
	title = {Entropy Production for Quantum Dynamical Semigroups},
	volume = {19},
	year = {1978}}

@article{Alicki1979,
	author = {Alicki, Robert},
	journal = {J. Phys. A},
	pages = {L103},
	title = {The Quantum Open System as a Model of the Heat Engine},
	volume = {12},
	year = {1979}}

@book{BreuerPetruccione,
	author = {Breuer, Heinz-Peter and Petruccione, Francesco},
	publisher = {Oxford University Press},
	title = {The Theory of Open Quantum Systems},
	year = {2002}}

@article{Jarzynski1997,
	author = {Jarzynski, Christopher},
	journal = {Phys. Rev. Lett.},
	number = {14},
	pages = {2690--2693},
	title = {Nonequilibrium Equality for Free Energy Differences},
	volume = {78},
	year = {1997}}

@article{Crooks1999,
	author = {Crooks, Gavin E.},
	journal = {Phys. Rev. E},
	number = {3},
	pages = {2721--2726},
	title = {Entropy Production Fluctuation Theorem and the Nonequilibrium Work Relation},
	volume = {60},
	year = {1999}}

@article{Seifert2012,
	author = {Seifert, Udo},
	journal = {Rep. Prog. Phys.},
	pages = {126001},
	title = {Stochastic Thermodynamics, Fluctuation Theorems and Molecular Machines},
	volume = {75},
	year = {2012}}

@article{Weinhold1975,
	author = {Weinhold, Frank},
	journal = {J. Chem. Phys.},
	pages = {2479--2483},
	title = {Metric Geometry of Equilibrium Thermodynamics},
	volume = {63},
	year = {1975}}

@article{Ruppeiner1995,
	author = {Ruppeiner, George},
	journal = {Rev. Mod. Phys.},
	pages = {605--659},
	title = {Riemannian Geometry in Thermodynamic Fluctuation Theory},
	volume = {67},
	year = {1995}}

@book{Hermann1973,
	author = {Hermann, Robert},
	publisher = {Marcel Dekker},
	title = {Geometry, Physics, and Systems},
	year = {1973}}

@article{Klatzow2019,
	author = {Klatzow, James and Becker, Jonas N. and Ledingham, Patrick M. and Weinzetl, Christian and Kaczmarek, Krzysztof T. and Saunders, Dylan J. and Nunn, Joshua and Walmsley, Ian A. and Uzdin, Raam and Poem, Eilon},
	doi = {10.1103/PhysRevLett.122.110601},
	journal = {Phys. Rev. Lett.},
	pages = {110601},
	title = {Experimental Demonstration of Quantum Effects in the Operation of Microscopic Heat Engines},
	volume = {122},
	year = {2019}}

@article{Camati2019,
	author = {Camati, Patrice A. and Santos, Jonas F. G. and Serra, Roberto M.},
	doi = {10.1103/PhysRevA.99.062103},
	journal = {Phys. Rev. A},
	pages = {062103},
	title = {Coherence effects in the performance of the quantum Otto heat engine},
	volume = {99},
	year = {2019}}

@article{Dorfman2018,
	author = {Dorfman, Konstantin E. and Xu, Dong and Cao, Jianshu},
	doi = {10.1103/PhysRevE.97.042120},
	journal = {Phys. Rev. E},
	pages = {042120},
	title = {Efficiency at maximum power of a laser quantum heat engine enhanced by noise-induced coherence},
	volume = {97},
	year = {2018}}

@article{Rahav2012,
	author = {Rahav, Saar and Harbola, Upendra and Mukamel, Shaul},
	doi = {10.1103/PhysRevA.86.043843},
	journal = {Phys. Rev. A},
	pages = {043843},
	title = {Heat fluctuations and coherences in a quantum heat engine},
	volume = {86},
	year = {2012}}

@article{Mrugala1991,
	author = {Mrugala, Ryszard and Nulton, James D. and Sch{\"o}n, J. Christian and Salamon, Peter},
	journal = {Rep. Math. Phys.},
	pages = {109--121},
	title = {Contact Structure in Thermodynamic Theory},
	volume = {29},
	year = {1991}}

@article{Avron2000,
	author = {Avron, J. E. and Fraas, M. and Graf, G. M. and Grech, P.},
	doi = {10.1007/s00220-008-0673-8},
	journal = {Communications in Mathematical Physics},
	pages = {651--675},
	title = {Adiabatic Response for Lindblad Dynamics},
	volume = {287},
	year = {2009}}

@article{Bravetti2017,
	author = {Bravetti, Alessandro},
	journal = {Entropy},
	pages = {535},
	title = {Contact Hamiltonian Dynamics: The Concept and Its Use},
	volume = {19},
	year = {2017}}

@article{Bravetti2015,
	author = {Bravetti, Alessandro and Lopez-Monsalvo, Cesar S. and Nettel, Francisco},
	journal = {Ann. Phys.},
	pages = {377--400},
	title = {Contact Symmetries and Hamiltonian Thermodynamics},
	volume = {361},
	year = {2015}}

@article{Scully2003,
	author = {Scully, Marlan O. and Zubairy, M. Suhail and Agarwal, Girish S. and Walther, Herbert},
	journal = {Science},
	pages = {862--864},
	title = {Extracting Work from a Single Heat Bath via Vanishing Quantum Coherence},
	volume = {299},
	year = {2003}}

@article{DeffnerLutz2011,
	author = {Deffner, Sebastian and Lutz, Eric},
	journal = {Phys. Rev. Lett.},
	pages = {140404},
	title = {Nonequilibrium Entropy Production for Open Quantum Systems},
	volume = {107},
	year = {2011}}

@article{Ren2010,
	author = {Ren, Jie and H{\"a}nggi, Peter and Li, Baowen},
	journal = {Phys. Rev. Lett.},
	pages = {170601},
	title = {Berry-Phase-Induced Heat Pumping},
	volume = {104},
	year = {2010}}

@article{Brandner2020,
	author = {Brandner, Kay and Saito, Keiji},
	journal = {Phys. Rev. Lett.},
	pages = {040602},
	title = {Thermodynamic Geometry of Microscopic Heat Engines},
	volume = {124},
	year = {2020}}

@article{Scandi2019,
	author = {Scandi, Matteo and Perarnau-Llobet, Mart{\'\i}},
	journal = {Quantum},
	pages = {197},
	title = {Thermodynamic Length in Open Quantum Systems},
	volume = {3},
	year = {2019}}

@article{Abiuso2020,
	author = {Abiuso, Paolo and Miller, Harry J. D. and Perarnau-Llobet, Mart{\'\i} and Scandi, Matteo},
	journal = {Entropy},
	pages = {1076},
	title = {Geometric Optimisation of Quantum Thermodynamic Processes},
	volume = {22},
	year = {2020}}

@article{TerrenAlonso2022,
	author = {Terr{\'e}n Alonso, Pablo and Abiuso, Paolo and Perarnau-Llobet, Mart{\'\i} and Arrachea, Liliana},
	journal = {PRX Quantum},
	pages = {010326},
	title = {Geometric Optimization of Nonequilibrium Thermal Machines},
	volume = {3},
	year = {2022}}

@article{Bhandari2020,
	author = {Bhandari, Bibek and Terr{\'e}n Alonso, Pablo and Taddei, Fabio and von Oppen, Felix and Fazio, Rosario and Arrachea, Liliana},
	journal = {Phys. Rev. B},
	pages = {155407},
	title = {Geometric Properties of Adiabatic Quantum Thermal Machines},
	volume = {102},
	year = {2020}}

@article{Zurek2003,
	author = {Zurek, Wojciech H.},
	journal = {Rev. Mod. Phys.},
	pages = {715--775},
	title = {Decoherence, Einselection, and the Quantum Origins of the Classical},
	volume = {75},
	year = {2003}}

\end{document}